\documentclass[11pt,a4paper]{article}
\usepackage{jheppub}
\usepackage{amscd}
\usepackage{amsfonts}
\usepackage{amsthm}
\usepackage{stmaryrd}
\usepackage{tikz}
\usetikzlibrary{arrows,shapes}

\def\IR{\mathbb{R}}
\def\ID{\mathbb{D}}

\newcommand\cmp[3]{{\it Commun.\ Math.\ Phys.\ }{\bf #1} (#2) #3}

\newcommand\npb[3]{{\it Nucl.\ Phys.\ }{\bf B #1} (#2) #3}
\newcommand\plb[3]{{\it Phys.\ Lett.\ }{\bf B #1} (#2) #3}

\newcommand\hepth[1]{{\tt hep-th/}#1}

\newcommand{\pp}{{=\!\!\!|}}
\newcommand\fverb{\setbox\pippobox=\hbox\bgroup\verb}
\newcommand\fverbdo{\egroup\medskip\noindent%
            \fbox{\unhbox\pippobox}\ }
\newcommand\fverbit{\egroup\item[\fbox{\unhbox\pippobox}]}
\newbox\pippobox

\title{A Note on Supersymmetric Chiral Bosons}

\author[a,b]{Alexander Sevrin}
\author[a]{ and Daniel C. Thompson}
\affiliation[a]{
Theoretische Natuurkunde, Vrije Universiteit Brussel\\ and\\The International Solvay Institutes\\
Pleinlaan 2, B-1050 Brussels, Belgium }
\affiliation[b]{also at the Physics Department, Universiteit Antwerpen\\
Campus Groenenborger, 2020 Antwerpen, Belgium}

\emailAdd{Alexandre.Sevrin@vub.ac.be}
\emailAdd{Daniel.Thompson@vub.ac.be}

\abstract{In this note we extend the Pasti-Sorokin-Tonin formalism for chiral bosons in two dimensions to  $N=(1,1)$
and $N=(2,2)$ superspace. In the latter case the formalism is developed for chiral, twisted chiral and semi-chiral superfields.}

\keywords{Superspace, sigma models, chiral bosons}

\begin{document}
 
\maketitle

\setcounter{equation}{0}

%
%

\section{Introduction} \label{introduction}
Chiral $p$-form fields may be defined for any even $p$ in a $D=2(p+1)$ dimensional Minkowski spacetime as potentials with self-dual $(p+1)$-form field strengths.  They are ubiquitous in supergravity and string theory; notable examples being the self-dual Ramond-Ramond five-form in the type IIB theory and the self-dual three-form in the $(2,0)$ tensor multiplet on the world volume of an $M5$ brane.  A prototype is found in the two-dimensional chiral boson which plays a prominent r\^ole in heterotic string constructions \cite{Gross:1984dd} and in the (T-duality symmetric) doubled formalism of string theory \cite{Tseytlin:1990nb,Tseytlin:1990va,Hull:2004in}. 

A particular challenge with these chiral fields is to incorporate the self-duality condition at the level of an action whilst at the same time preserving manifest Lorentz invariance.   The difficulties arise primarily because self-duality is a first order differential condition and gives rise to second class constraints.   A Lorentz invariant formulation for chiral $p$-forms was developed by Pasti, Sorokin and Tonin \cite{Pasti:1996vs}.  The essence of this approach is to introduce auxiliary fields that furnish the theory with a gauge invariance.  This  invariance serves to render the anti-self-dual part of the field strength pure gauge. 

It is  instructive to look at the description of the $N=(1,0)$ supersymmetric chiral boson. A Siegel type formulation is obtained by considering a scalar field coupled to $N=(1,0)$ supergravity in the lightcone gauge \cite{Siegel:1983es}. As can be seen from {\em e.g.} \cite{Grisaru:1987pf,HariDass:1988dp}, this approach is rather cumbersome. A more immediate way -- leading to the same result --  starts from the scalar field theory and proceeds by gauging a chiral conformal symmetry. Indeed, consider the action,
\begin{eqnarray}
{\cal S}=2i\int d^2 \sigma \,d \theta^+\,
D_+ \phi\, \partial _= \phi\,,
\end{eqnarray}
which is invariant under,
\begin{eqnarray}
\delta \phi= \varepsilon^= \partial _= \phi\,,
\end{eqnarray}
with $D_+ \varepsilon^==0$. Gauging the symmetry is straightforward and amounts to minimal coupling. Introducing a gauge field $h_{\pp\,+}$ we obtain the action,
\begin{eqnarray}
{\cal S}=2i\int d^2 \sigma \,d \theta^+\left(
D_+ \phi \partial _= \phi- h_{\pp\, +}\partial _= \phi\, \partial _= \phi
\right)\,,
\end{eqnarray}
which is invariant under the chiral superconformal gauge transformations,
\begin{eqnarray}
\delta h_{\pp\, +}&=& D_+ \varepsilon^=+ \varepsilon ^= \partial _=h_{\pp\, +} - \partial _= \varepsilon^=\, h_{\pp\, +}\,, \nonumber\\
\delta \phi&=& \varepsilon^= \partial _= \phi\,,
\end{eqnarray} 
with $\varepsilon^= = \varepsilon^=(\sigma^\pp, \sigma ^=, \theta^+)$. Passing to the Floreanini-Jackiw gauge\footnote{This gauge choice is manifestly not Lorentz invariant but neither is the Floreanini-Jackiw action.} $h_{\pp\,+}=-i\, \theta^+/2$, the action becomes,
\begin{eqnarray}
{\cal S}=2i\,\int d^2 \sigma \, d \theta^+
\big(\partial _+ \phi  - \frac i 2 \theta^+ \partial _ \sigma \phi\big)\, \partial _= \phi\,.\label{FJ10}
\end{eqnarray}
The equations of motion for $\phi$ are solved by $\phi= \varphi( \sigma^\pp, \theta^+)+f( \tau)$ where the $f( \tau )$ factor can be eliminated using the fact that the action is invariant under the {\em gauge transformation} $ \phi \rightarrow \phi- f( \tau)$. Despite the fact that $f$ only depends on $\tau$ and not on $\sigma$, it is truly a gauge symmetry in the sense of  \cite{MHbook} (see also \cite{ Henneaux:1987hz,Bekaert:1998yp}): a gauge symmetry corresponds to a symmetry with a vanishing Noether charge. In the present case it is not hard to verify that the Noether charge corresponding to the symmetry $\delta \phi= -f(\tau)$ is indeed zero. So putting $\phi= \varphi( \sigma^\pp, \theta^+)$ is a gauge choice. This has to be contrasted with the symmetry of the action $\delta \phi= \varphi( \sigma^\pp\,,\theta^+)$ which has a non-vanishing Noether charge and which gives rise to rigid symmetries of the action of the Kac-Moody type. An alternative point of view adopted in much of the literature is to  simply invoke boundary conditions which enforce  $\phi= \varphi( \sigma^\pp, \theta^+) $   however in the context of closed strings   such an approach is inappropriate.  

Concluding, the action eq.~(\ref{FJ10})  does describe an $N=(1,0)$ supersymmetric chiral boson.  Expanding into components one finds that, with this gauge choice, the action consists of a chiral boson with a Floreanini-Jackiw action  \cite{Floreanini:1987as} whose supersymmetric completion is a single Majorana-Weyl fermion. 

An alternative formulation -- which will turn out to be most useful for the $N=(2,2)$ supersymmetric case -- parameterizes the gauge field by a single bosonic superfield $F^=(\sigma^\pp, \sigma ^=, \theta^+)$,
\begin{eqnarray}
h_{\pp\,+} \partial _==\big(D_+\,e^{F^= \partial _=}\big)\,e^{-F^= \partial _=}\,,
\end{eqnarray}
so that the action is now given by,
\begin{eqnarray}
{\cal S}=2i\,\int d^2 \sigma \, d \theta^+\,
D_+\big(e^{-F^= \partial _=}\phi\big)
\partial _=\big(e^{-F^= \partial _=}\phi\big)
\,.\label{FJ1000}
\end{eqnarray}
It is invariant under the gauge transformations,
\begin{eqnarray}
\phi \rightarrow e^{F^= \partial _=}\phi\,,\qquad e^{F^= \partial _=} \rightarrow e^{\varepsilon^= \partial _=}e^{F^= \partial _=}e^{\zeta^= \partial _=}\,,\label{trtr5}
\end{eqnarray}
where $\varepsilon ^=$ is an arbitrary superfield and $\zeta^=$ satisfies $D_+\zeta^==0$.
Upon varying eq.~(\ref{FJ1000})  we get,
\begin{eqnarray}
\delta {\cal S}&=&4i\,\int d^2 \sigma \, d \theta^+\,\big(\delta e^{F^= \partial _=}e^{-F^= \partial _=}\phi\big) 
\partial _=\big(e^{F^= \partial _=}D_+( e^{-F^= \partial _=}\phi)\big)\nonumber\\
&&-4i\,\int d^2 \sigma \, d \theta^+\,\delta \phi\,\partial _=\big(e^{F^= \partial _=}D_+( e^{-F^= \partial _=}\phi)\big)\,.
\end{eqnarray}
From this one sees that the equation of motion for the gauge field is satisfied once $\phi $ satisfies its equation of motion whose 
solution is given by,
\begin{eqnarray}
\phi= \varphi(\sigma^\pp,\theta^+)+ e^{F^= \partial _=}\,f( \sigma^=)\,.
\end{eqnarray}
As the action is  invariant under the gauge transformations, 
\begin{eqnarray}
\phi \rightarrow \phi-e^{F^= \partial _=}\,f( \sigma^=)\,,\qquad F^= \rightarrow  F^=\,,
\end{eqnarray}
one recovers a chiral boson $\phi$ satisfying $ \partial _= \phi=0$. Alternatively we could also have made the gauge choice $F^==\sigma^\pp$ which brings us back to the Floreanini-Jackiw gauge.

The previous is very closely related to the formalism proposed by Pasti, Sorokin and Tonin \cite{Pasti:1995us}. In this approach a Beltrami-like parameterization is   used for the super zweibein in the lightcone gauge. It introduces a bosonic superfield $g^=$ which is related to the previous by\footnote{Note
that instead of $\sigma^=$ we could have written any function of $\sigma ^=$ which just reflects the symmetry transformation parameterized by $\zeta^=$ in eq.~(\ref{trtr5}).},
\begin{eqnarray}
g^==e^{F^= \partial _=}\,  \sigma^= e^{-F^= \partial _=}\,.\label{belp}
\end{eqnarray}
Acting with $D_+$ on eq.~(\ref{belp}) immediately results in,
\begin{eqnarray}
h_{\pp\,+} \partial _==\big(D_+\,e^{F^= \partial _=}\big)\,e^{-F^= \partial _=}= \frac{D_+ g^=}{\partial _=g^=}\, \partial _=\,.\label{1113}
\end{eqnarray}
With this one obtains the PST-action for the $N=(1,0)$ supersymmetric chiral boson,
\begin{eqnarray}
{\cal S}=2i\int d^2 \sigma \,d \theta^+\left(
D_+ \phi\, \partial _= \phi- \frac{D_+g^=}{ \partial _=g^=} \partial _= \phi\, \partial _= \phi
\right)\,,\label{action10}
\end{eqnarray}
which is invariant under the gauge transformations,
\begin{eqnarray}
\delta g^== \varepsilon^= \partial _= g^=\,, \qquad \delta \phi= \varepsilon^= \partial _= \phi\,.
\end{eqnarray}
Identifying $\xi^=\equiv \varepsilon^= \partial _= g^=$, we obtain the PST gauge transformations,
\begin{eqnarray}
\delta \phi= \xi^=\,\frac{ \partial _= \phi}{ \partial _= g}\,,\qquad
\delta g= \xi^=\,,\label{trsf10}
\end{eqnarray}
where 
\begin{eqnarray}
\xi^== \xi^=( \sigma^=, \sigma ^\pp, \theta^+)\,.
\end{eqnarray}
In this parameterization the FJ gauge choice corresponds\footnote{Note that this choice is not unique as the expression for $h_{\pp\,+}$ in eq.~(\ref{1113}) is invariant under $g^=\rightarrow g'^=(g^=)$. So we could as well have taken $g^==g^=(\tau)$.} to taking $g^== 2\tau $. 

Let us now compare the Siegel and the PST formulations. In the Siegel approach, the equation of motion for $h_{\pp\,+}$ becomes,
\begin{eqnarray}
\partial _=\phi \,\partial _= \phi=0\,,
\end{eqnarray}
while in the PST approach we get for the equation of motion for $g^=$,
\begin{eqnarray}
\partial _= \phi\partial _=\big(D_+ \phi - \frac{D_+g^=}{\partial _= g^=}\partial _= \phi\big)=0\,.
\end{eqnarray}
One now clearly sees the difference between the PST and the Siegel formulation.
Indeed the equation of motion for $\phi $ implies the equation of motion for $g^=$ in the PST formulation while this is not the case in the Siegel formulation where
the equation of motion for $h_{\pp\,+}$ survives -- after choosing a gauge -- as a constraint. In  \cite{Pasti:1996vs} it was shown that from the classical Hamiltonian perspective the PST action has a first class primary constraint associated to the gauge symmetry which is not simply the square of the second class chirality constraint as is the case in the Siegel formalism.   

In this paper we will extend the PST formulation to a supersymmetric chiral boson in a manifestly super-Poincar\'e invariant way. 
Our strategy starts with the construction of the appropriate supergravity theory in the light-cone gauge which we obtain by gauging a chiral superconformal transformation. Subsequently we rewrite the gauge field(s) in a Beltrami like way and identify the PST gauge transformations. In the next section we will develop an $N=(1,1)$ superspace formalism which is followed in the third section by an $N=(2,2)$ superspace description of chiral bosons. We end with some conclusions.
 The first appendix  summarizes our conventions, appendices B and C provide a brief reviews of superconformal transformations in $N=(1,1)$ and $N=(2,2)$ superspace respectively.  In what follows we make intensive use of the information contained in these appendices. 

\section{$N=(1,1)$ supersymmetric chiral bosons} 
The action for a free scalar in $N=(1,1)$ superspace,
\begin{eqnarray}
{\cal S}=4\int d^2 \sigma d^2 \theta \, D_+ \phi D_- \phi\, ,
\end{eqnarray}
is invariant under the superconformal transformation,
\begin{eqnarray}
\phi \rightarrow \phi'= e^{{\cal O}(\varepsilon^=)}\, \phi\,,
\end{eqnarray}
provided $D_+ \varepsilon^==0$. The operator ${\cal O}(\varepsilon^=)$ entering in the above may be ascertained from the infinitesimal variation, 
\begin{eqnarray}
\label{deltaphi11}
\delta \phi=  {{\cal O}(\varepsilon^=)} \phi \equiv \varepsilon^= \partial _= \phi+iD_- \varepsilon^= D_- \phi\,.
\end{eqnarray}
In order to promote this to a full gauge symmetry we introduce a light-cone zweibein $h_{\pp \,+}$ transforming as,
\begin{eqnarray}
\delta h_{\pp\, +}= D_+ \varepsilon^=+ \varepsilon ^= \partial _= h_{\pp\, +}+i D_- \varepsilon^=D_-h_{\pp\, +}- \partial_= \varepsilon^=h_{\pp\, +}\,,\label{n11gf}
\end{eqnarray}
and we replace the derivative $D_+ \phi$ by the covariant derivative $\nabla_+ \phi $,
\begin{eqnarray}
\nabla_+ \phi= D_+ \phi -h_{\pp\, +} \partial _= \phi+i D_-h_{\pp\, +}D_- \phi\, ,\label{covdev11}
\end{eqnarray}
which results in the gauge invariant action,
\begin{eqnarray}
{\cal S}&=&4\int d^2 \sigma d^2 \theta \, \nabla_+ \phi D_- \phi \nonumber\\
&=&
4\int d^2 \sigma d^2 \theta \, \left( D_+ \phi D_- \phi 
-h_{\pp\, +}\,\partial _= \phi D_- \phi 
\right)\,.
\label{n11gauged}
\end{eqnarray}
This yields a Siegel type formulation of the supersymmetric chiral boson. 

An alternative approach introduces a superfield $F^=$ in terms of which the covariant derivative is given by,
\begin{eqnarray}
\nabla_+\phi=D_+\phi -\big(D_+e^{{\cal O}(F^=)}\big)e^{-{\cal O}(F^=)}\phi=e^{{\cal O}(F^=)}D_+\big(e^{-{\cal O}(F^=)}\phi\big)\,,
\end{eqnarray}
where $F^=$ transforms under a gauge transformation as,
\begin{eqnarray}
e^{{\cal O}(F^=)} \rightarrow e^{{\cal O}(F'^=)}=e^{{\cal O}(\varepsilon^=)}e^{{\cal O}(F^=)}e^{{\cal O}(\zeta^=)}\,,
\end{eqnarray}
where $D_+ \zeta^==0$. With this one rewrites eq.~(\ref{n11gauged}) as,
\begin{eqnarray}
{\cal S}&=&4\int d^2 \sigma d^2 \theta \, \Big(e^{{\cal O}(F^=)} D_+ (e^{-{\cal O}(F^=)}\phi)\Big) D_- \phi \nonumber\\
&=&4\int d^2 \sigma d^2 \theta \,  D_+\big( e^{-{\cal O}(F^=)}\phi\big) D_-\big(e^{-{\cal O}(F^=)} \phi\big) \,.
\end{eqnarray} 
From the variation of the action,
\begin{eqnarray}
\delta{\cal S}&=&8\,\int d^2 \sigma d^2 \theta \, \delta \phi\,D_-\big(e^{{\cal O}(F^=)} D_+( e^{-{\cal O}(F^=)}\phi)\big)\nonumber\\
&&+8\,\int d^2 \sigma d^2 \theta \,
\big(\delta e^{{\cal O}(F^=)}e^{-{\cal O}(F^=)}\phi\big)\,D_-\big(e^{{\cal O}(F^=)} D_+( e^{-{\cal O}(F^=)}\phi)\big)\,,
\end{eqnarray}
one finds that once again the equation of motion for the gauge field is satisfied when $\phi $ is on-shell and $\phi$ is then given by,
\begin{eqnarray}
\phi=\varphi(\sigma^\pp, \theta^+)+ e^{{\cal O}(F^=)}f(\sigma^=, \theta^-)\,,
\end{eqnarray}
where $\varphi$ and $f$ satisfy $D_- \varphi =0$ and $D_+ f=0$. The action is invariant under the gauge symmetry $\phi \rightarrow \phi-e^{{\cal O}(F^=)}f(\sigma^=, \theta^-)$ which shows that we indeed describe an $N=(1,1)$ supersymmetric chiral boson. 

In order to pass to a PST parameterization we introduce the Beltrami parameterization in terms of a bosonic superfield $g^=$ and a fermionic one $\psi^-$,
\begin{eqnarray}
g^=&=&e^{{\cal O}(F^=)}\sigma^= e^{-{\cal O}(F^=)}\,,\nonumber\\
\psi^-&=&e^{{\cal O}(F^=)}\theta ^- e^{-{\cal O}(F^=)}\,.\label{defBel2}
\end{eqnarray}
These fields are clearly not independent and using the results of appendix B one shows that they satisfy the constraint,
\begin{eqnarray}
D_- g^=+\frac i 2 \psi^- D_- \psi^-=0 \,,\label{n11con}
\end{eqnarray}
which implies,
\begin{eqnarray}
\label{n11con2}
\partial _=g^=-\frac i 2 \psi^- \,\partial _= \psi^-= D_- \psi^- D_- \psi^-\,.
\end{eqnarray}
By construction they transform as superconformal primary fields under the gauge transformation,
\begin{eqnarray}
\delta g^== \varepsilon^= \partial _= g^=+iD_- \varepsilon^= D_- g^=\,,\qquad
\delta \psi^-= \varepsilon^= \partial _= \psi^-+iD_- \varepsilon^= D_- \psi^-\,.\label{n11trsf}
\end{eqnarray}
Using the defining relations eq.~(\ref{defBel2}) we get,
\begin{eqnarray}
D_+g^=={\cal O}(h_{\pp\,+})g^=\,,\qquad
D_+\psi^-={\cal O}(h_{\pp\,+})\psi ^-\,,
\end{eqnarray} 
from which one obtains (making use of eqs.~(\ref{n11con}) and (\ref{n11con2})),
\begin{eqnarray}
h_{\pp \, +}= \frac{D_+ g^=+\frac i 2 \psi^- D_+ \psi ^-}{D_- \psi^- D_- \psi^- }\,,\label{hBel2}
\end{eqnarray}
and one may readily  verify  that eqs.~(\ref{n11trsf}) and (\ref{n11con}) indeed imply eq.~(\ref{n11gf}).   
With this we finally arrive at the PST action in $N=(1,1)$ superspace,
\begin{eqnarray}
{\cal S}=4\int d^2 \sigma d^2 \theta \, \left( D_+ \phi D_- \phi 
-\frac{D_+ g^=+\frac i 2 \psi^- D_+ \psi^- }{D_- \psi^- D_- \psi^- }\,\partial _= \phi D_- \phi 
\right)\,,\label{210}
\end{eqnarray}
which is invariant under the PST gauge transformations,
\begin{eqnarray}
&&\delta g^== \xi^=+\frac 1 2 \, D_-\xi^=\frac{\psi^-}{D_- \psi^-} \,, \qquad \delta \psi^-=i\frac{D_- \xi^=}{D_- \psi ^-}\,, \nonumber\\
&&\delta \phi=\frac{\xi^=}{\big(D_-\psi^-\big)^2}\, \partial _= \phi+i D_-\left(
\frac{\xi^=}{\big(D_-\psi^-\big)^2}
\right)\,D_- \phi \,,\label{pst11}
\end{eqnarray}
where the transformation parameter $\xi^=$ is an arbitrary real infinitesimal unconstrained $N=(1,1)$ superfield. These transformations may be obtained from the left-handed local superconformal transformations of eqs.~(\ref{deltaphi11})  and (\ref{n11trsf}) with the gauge parameter redefined according to $\xi^= = \epsilon^= D_- \psi^- D_- \psi^- $. Once again the
transformations eq.~(\ref{pst11}) are compatible with the constraints eq.~(\ref{n11con}).

Choosing $F^==\sigma^\pp$ we reach the Floreanini-Jackiw gauge. Using eq.~(\ref{defBel2}) one finds that this is equivalent to $g^==2\tau$ and $\psi^-=\theta^-$ which results in,
\begin{eqnarray}
h_{\pp\,+}=-\frac i 2\,\theta ^+\,.\label{FJ11}
\end{eqnarray}
The action is simply,
\begin{eqnarray}
{\cal S}=4\int d^2 \sigma d^2 \theta \, \big( \partial _+ -\frac i 2 \, \theta^+ \partial _\sigma\big)\phi\, D_-\phi \,,\label{bbvv}
\end{eqnarray}
and the solution to the equation of motion is given by,
\begin{eqnarray}
\phi= \varphi(\sigma^\pp,\theta^+)+f(\tau,\theta^-)\,,
\end{eqnarray}
where $D_ -\varphi=0$ and where the last term can be eliminated using the gauge symmetry $\phi \rightarrow \phi-f(\tau,\theta^-)$. So we end up with a supersymmetric chiral boson satisfying $D_-\phi=0$ and as a consequence $\partial _=\phi=0$ holds as well.
 
\section{$N=(2,2)$ supersymmetric chiral bosons}
\subsection{Chiral superconformal gauge theories}
In order to obtain the PST action for chiral bosons in $N=(2,2)$ superspace we could have started from $d=2$, $N=(2,2)$ supergravity 
in the light-cone gauge as developed in \cite{Grisaru:1995kn}. However as that approach turns out to be quite involved we start by gauging the left-handed $N=(2,2)$ superconformal symmetry  and obtain in this way the PST formulation.\footnote{Chiral bosons in $N=(2,2)$ superspace were also considered in \cite{Bellucci:1991id}  (predating the PST formalism which we are concerned with). }

An infinitesimal superconformal transformation of an $N=(2,2)$ superfield $\Phi$ with conformal weight $\Delta$ and $U(1)$ charge $q$ is given by,
\begin{eqnarray}
\delta \Phi=  \frac i 2 \, \ID_- \big(\varepsilon^=\bar\ID_- \Phi\big)+ \frac i 2 \,\bar  \ID_- \big(\varepsilon^=\ID_- \Phi\big)+
\Delta\, \partial _= \varepsilon^= \, \Phi+i\,q\,\big([\ID_-,\bar\ID_-] \varepsilon^=\big)\,\Phi\,,\label{suco22}
\end{eqnarray}
where $\varepsilon^= $ is a real with $\ID_+ \varepsilon^==\bar\ID_+\varepsilon^= =0$ (and consequently $\partial _\pp\, \varepsilon =0$ as well) and which satisfies the algebra,
\begin{eqnarray}
\big[\delta( \varepsilon^=_1),\delta( \varepsilon^=_2)\big]\,\Phi=\delta(
\varepsilon_2^= \partial _= \varepsilon_1^=- \partial _= \varepsilon_2^= \varepsilon _1^=
+\frac i 2\, \ID_- \varepsilon_2^=\bar \ID_- \varepsilon _1^=+\frac i 2\,\bar \ID_- \varepsilon_2^= \ID_- \varepsilon _1^=
)\, \Phi\,.
\end{eqnarray}
From now we will exclusively deal with neutral scalar fields, {\em i.e.} $\Delta=q=0$ in eq.~(\ref{suco22}) . 
With this we can write a superconformal transformation as
\begin{eqnarray}
\Phi \rightarrow \Phi'= e^{ {\cal O}( \varepsilon^=)}\Phi\,,\label{suco222}
\end{eqnarray}
where
\begin{eqnarray}
{\cal O}( \varepsilon^=) &=& \frac{i}{2}\, \ID_-\big( \varepsilon^=\bar\ID_-  +   \frac{i}{2}\, \bar \ID_- \big(\varepsilon^= \ID_-   \,,\nonumber\\
&=&   \varepsilon^= \partial_=  +\frac{i}{2} \,(\ID_- \varepsilon^=) \bar\ID_-  +   \frac{i}{2} \,(\bar \ID_- \varepsilon^=) \ID_-   \, . 
\end{eqnarray}
Here we are using the somewhat standard notation that an unclosed parenthesis indicates an operator acting on everything to its right. 

We now gauge the symmetry using the maximal approach, {\em i.e.} we take $\varepsilon^=$ to be a complex unconstrained superfield. Whenever a superfield $\Phi$ satisfies the constraints $\ID_-\Phi=0$ or 
$\bar\ID_-\Phi=0$ then so does $\Phi'$ (where $\Phi'$ was defined in eq.~(\ref{suco222})). However this is no longer so for constraints of the form
$\ID_+\Phi=0$ or $\bar\ID_+\Phi=0$. In that case we have to replace the ordinary derivatives by covariant derivatives,
\begin{eqnarray}
\nabla_+=\ID_+- \ID_+e^ {{\cal O}(V^=)}e^{-{\cal O}(V^=)}\,,\qquad
\bar\nabla_+=\bar \ID_+- \bar\ID_+e^ {{\cal O}(W^=)}e^{-{\cal O}(W^=)}\,,
\end{eqnarray}
where $V^=$ and $W^=$ are complex unconstrained superfields. They transform as,
\begin{eqnarray}
e^{{\cal O}(V^=)} &\rightarrow& e^{{\cal O}(V'^=)}= e^{{\cal O}(\varepsilon^=)}e^{{\cal O}(V^=)}e^{-{\cal O}(\xi^=)}\,, \nonumber\\
e^{{\cal O}(W^=)} &\rightarrow& e^{{\cal O}(W'^=)}= e^{{\cal O}(\varepsilon^=)}e^{{\cal O}(W^=)}e^{-{\cal O}(\eta^=)}\,,
\end{eqnarray}
where $\xi^=$ and $\eta^=$ are complex superfields satisfying $\ID_+\xi^==\bar\ID_+\eta^==0$. We
proceed by making the gauge choice $W^==0$ and imposing the conventional constraint ${\cal F}_{\pp}=0$ where,
\begin{eqnarray}
\big\{\nabla_+,\bar\nabla_+\big\}=-2i\,\nabla_\pp- {\cal O}({\cal F}_{\pp}) \,,
\end{eqnarray}
and $\nabla_\pp= \partial _\pp- {\cal O}(A_\pp)$. This constraint can be solved for $A_\pp\,$. Finally we require that $A_\pp^\dagger=A_\pp\,$ holds modulo a gauge transformation. This gives $V^={}^\dagger=-V^=$.
Introducing $F^=$, $F^==i\,V^=$ we arrive at,
\begin{eqnarray}
\nabla_+=\ID_++ e^ {-i\,{\cal O}(F^=)}\ID_+e^{i\,{\cal O}(F^=)}\,,\qquad
\bar\nabla_+=\bar \ID_+\,.
\end{eqnarray}
 As detailed in the Appendix C, this covariant derivative may be recast as 
\begin{eqnarray}
\label{nablawithh}
\nabla_+  =  \ID_+ - h_{\pp\, +}   \partial_= + \frac{i}{2} \bar\ID_-   h_{\pp\, +}   \ID_- + \frac{i}{2}  \ID_-   h_{\pp\, +}  \bar \ID_-  \, , 
\end{eqnarray}
with $h_{\pp\, +}$ and $F^=$  related by an $N=(2,2)$  Beltrami like parametrization. 
The residual gauge transformations are given by,
\begin{eqnarray}
e^{i{\cal O}(\,F^=)} &\rightarrow& e^{i{\cal O}( F'^=)}= e^{{\cal O}(\bar\varepsilon^=)}e^{i{\cal O}(F^=)}e^{-{\cal O}(\varepsilon ^=)}\,\nonumber \,, \\
\Phi& \rightarrow& \Phi' =e^{{\cal O}(\varepsilon^=)}\Phi\,,
\end{eqnarray}
where $\bar \varepsilon^= =\varepsilon^{=\dagger}$ and $\bar\ID_+ \varepsilon ^==\ID_+ \bar\varepsilon^==0$.  
We now consider infinitesimal gauge transformations and we get from the above,
\begin{eqnarray}
e^{-i {\cal O}(F^=)} \delta e^{i {\cal O}(F^=)}= e^{-i {\cal O}(F^=)}  {\cal O}\big( \bar \varepsilon^=\big)
e^{i {\cal O}(F^=)} -{\cal O}\big( \varepsilon^=\big)\,.
\end{eqnarray}
From this one obtains the variation of $F^=$ itself order by order in $F^=$ using a Baker-Campbell-Hausdorff expansion
\begin{eqnarray}
&&\delta F^==i\big(\varepsilon^=-\bar \varepsilon ^=\big)+\frac 1 2\,(\varepsilon^=+\bar\varepsilon^=) \partial _=F^=+
\frac i 2\,\bar\ID_-\big( \frac 1 2\,(\varepsilon^=+\bar\varepsilon^=) \big)\,\ID_-F^=+ \nonumber\\
&&\quad\frac i 2\,\ID_-\big( \frac 1 2\,(\varepsilon^=+\bar\varepsilon^=) \big)\,\bar\ID_-F^=-
\partial _=\big( \frac 1 2\,(\varepsilon^=+\bar\varepsilon^=) \big)\,F^=+\mbox{ order $F^2$ and higher}\,.\label{trFBCH}
\end{eqnarray}
 One can see that the imaginary part of $\varepsilon^{=}$ gives a shift, which one could employ to reach Wess-Zumino gauge, and at first order the real part of $\varepsilon^{=}$ gives a superconformal transformation of $\Delta=-1$ and $q=0$.  
Indeed making a partial gauge choice such that $F^=\big|_{\theta^+=\bar \theta ^+=0}=0$ results in $\mbox{Im}(\varepsilon^=)\big|_{\theta^+=\bar \theta ^+=0}=0$ and one verifies that the order $F^2$ and higher terms in eq.~(\ref{trFBCH})  vanish (for this one uses the observation that odd powers of $F^=$ in the rhs of eq.~(\ref{trFBCH}) transform with $\varepsilon^=+\bar \varepsilon ^=$ while even powers of $F^=$ go with $i(\varepsilon^=-\bar \varepsilon ^=)$). 
Note that  there is enough symmetry to complete gauge away this field (see  \cite{Grisaru:1995kn} for a discussion of this); this is entirely  what is needed since we did not wish to introduce any extra degrees of freedom, only auxiliary fields. 

Finally we turn to the gauge choice where the $N=(1,1)$ gauge multiplet is recovered. Indeed, choosing,
\begin{eqnarray}
F^==-2\hat \theta^+h_{\pp\,+}\,\label{Fgc11}
\end{eqnarray}
where $h_{\pp\,+}$ is independent of $\hat \theta ^-$, leaves a residual gauge symmetry parameterized by $\varepsilon^==\varepsilon _0+i\hat \theta^+D_+ \varepsilon_0$ and $\bar\varepsilon^==\varepsilon _0-i\hat \theta^+D_+ \varepsilon_0$ where $\varepsilon _0^\dagger= \varepsilon_0$
and where $\varepsilon _0$ does not depend on either $\hat \theta ^+$ or $\hat \theta^-$. One then verifies that eq.~(\ref{trFBCH}) agrees with eq.~(\ref{n11gf})  where $\varepsilon^=$ in eq.~(\ref{n11gf}) is identified with $\varepsilon_0$. As a consequence we can now -- using eq.~(\ref{Fgc11}) and eq.~(\ref{FJ11}) -- identify the Floreanini-Jackiw gauge in $N=(2,2)$ superspace:
\begin{eqnarray}
F^==2\,\theta^+\bar \theta ^+\,.
\end{eqnarray}

In summary, one may perform a chiral gauging of the $N=(2,2)$ supersymmetry by keeping the action unaltered but replacing any constraint of the form $\ID_+ \Phi=0$ with,
\begin{eqnarray}
0=\nabla_+\Phi=e^ {-i\,{\cal O}(F^=)}\ID_+\big(e^{i\,{\cal O}(F^=)}\Phi\big)\,,
\end{eqnarray}
or alternatively, by keeping the constraint $\ID_+ \Phi=0$ unaltered but replacing all occurrences of $\Phi$ in the action by $e^{-i\,{\cal O}(F^=)}\Phi\,$.   

In $N=(2,2)$ superspace scalar fields come into three flavors 
satisfying constraints linear in the derivatives\footnote{Note that this 
list is not exhaustive. Indeed when allowing for constraints quadratic in the derivatives one also gets complex linear and twisted complex linear superfields. They are dual to chiral and twisted chiral fields resp. Real versions of those fields exist as well. Chiral, twisted chiral and semi-chiral superfields are sufficient to give a manifest off-shell $N=(2,2)$ supersymmetric description of any $d=2$ $N=(2,2)$ supersymmetric non-linear-model \cite{Lindstrom:2005zr,Sevrin:1996jr}. The corresponding geometry is precisely generalized K\"ahler geometry proposed by Hitchin  \cite{Hitchin:2004ut} and developed by Gualtieri \cite{Gualtieri:2003dx}. The extra right- and lefthanded supersymmetry each define a complex structure which we call $J_+$ and $J_-$. Decomposing the tangent space to the target manifold as $\ker (J_++J_-)\oplus \ker(J_+-J_-)\oplus \mbox{im}({[}J_+,J_-{]}g^{-1})$, one finds that the
three subspaces are parameterized by chiral, twisted chiral and semi-chiral coordinates resp.}.
\begin{description}
 \item[Chiral superfields:] $z$, $\bar z$, obeying 
\begin{eqnarray}
 \bar \ID_\pm z =\ID_\pm \bar z=0\,.\label{cch}
\end{eqnarray} 
  \item[Twisted chiral superfields \cite{Gates:1984nk}:] $w$, $\bar w$,  obeying
\begin{eqnarray}
 \bar \ID_+w=\ID_-w=\ID_+\bar w=\bar \ID_-\bar w=0\,.\label{ctwch}
\end{eqnarray}
 \item[Semi-chiral superfields \cite{Buscher:1987uw}:] $l$, $\bar l$, $r$, $\bar r$,  obeying
\begin{eqnarray}
\bar \ID_+ l=\ID_+\bar l= \bar \ID_-r=\ID_-\bar r=0\,.\label{csch}
\end{eqnarray}
 \end{description}
Chiral and twisted chiral $N=(2,2)$ superfields have the same number of
component fields as $N=(1,1)$ superfields while semi-chiral $N=(2,2)$
superfields have twice as many, half of which are auxiliary  from the $N=(1,1)$
superspace point of view. In the next subsection we will first treat the simple case involving chiral or twisted chiral superfields. In the subsequent subsection we will investigate the semi-chiral case.

\subsection{Chiral and twisted chiral superfields}
The free field action for a chiral superfield is simply,
\begin{eqnarray}
{\cal S}_{\mbox{\footnotesize chiral}}=\int d^2 \sigma d^4 \theta\, z \bar z\,,
\end{eqnarray}
and it is invariant under the superconformal transformation eq.~(\ref{suco222}) (implementing the appropriate constraints eqs.~(\ref{cch})). Using the formalism developed in the previous subsection we get the gauged action,
\begin{eqnarray}
{\cal S}_{\mbox{\footnotesize chiral}}=\int d^2 \sigma d^4 \theta\, z\, e^{\frac 1 2\, \ID_-\big(F^=\,\bar\ID_-}\bar z\,, \label{gaugedchiral}
\end{eqnarray}
which is invariant under the gauge transformations,
\begin{eqnarray}
&& z \rightarrow z'= e^{{\cal O}(\varepsilon^=)}z\,,\qquad
\bar z \rightarrow \bar z'= e^{{\cal O}(\bar\varepsilon^=)}\bar z\,, \nonumber\\
&&e^{i{\cal O}(\,F^=)} \rightarrow e^{i{\cal O}( F'^=)}= e^{{\cal O}(\bar\varepsilon^=)}e^{i{\cal O}(F^=)}e^{-{\cal O}(\varepsilon ^=)}\,,
\end{eqnarray}
where $\bar \varepsilon^= =\varepsilon^{=\dagger}$ and $\bar\ID_+ \varepsilon ^==\ID_+ \bar\varepsilon^==0$.

A direct way of seeing that this indeed describes a chiral boson starts from the equation of motion for $z$,
\begin{eqnarray}
\ID_+\ID_-\Big(e^{-\frac 1 2 \, \bar\ID_-\big(F^=\ID_-}z\Big)=0\,,
\end{eqnarray}
 which is solved by,
\begin{eqnarray}
z=\hat z +e^{\frac 1 2 \, \bar\ID_-\big(F^=\ID_-}\ID_+ \Lambda\,,
\end{eqnarray}
where $\hat z$ satisfies $\bar\ID_\pm\hat z=\ID_- z=0$ and the second term is annihilated by both $\bar \ID_+$ and $\bar \ID_-$. It can be eliminated by a gauge transformation,
\begin{eqnarray}
z \rightarrow z -e^{\frac 1 2 \, \bar\ID_-\big(F^=\ID_-}\ID_+ \Lambda\,,\qquad F^= \rightarrow F^=\,,
\end{eqnarray}
leaving us with a chiral boson defined by $\bar\ID_\pm z=\ID_-z=0$ and $\ID_\pm \bar z=\bar\ID_-\bar z=0$ which obviously imply $\partial _=z= \partial _=\bar z=0$ as well.
 
Alternatively we could fix the gauge as in eq.~(\ref{Fgc11}) and pass to $N=(1,1)$ superspace,
\begin{eqnarray}
{\cal S}_{\mbox{\footnotesize chiral}}&=&\int d^2 \sigma d^4 \theta\, \big(z\bar z+\hat \theta^+h_{\pp\,+}\ID_-z\bar\ID_-\bar z\big) \nonumber\\
&=&4\int d^2 \sigma d^2 \theta\, \nabla_+z\,D_-\bar z\,,
\end{eqnarray}
where the covariant derivative was defined in  eq.~(\ref{covdev11}). The resulting action is precisely eq.~(\ref{n11gauged}) for a complex scalar clearly showing that we indeed describe a chiral complex boson.

For a twisted chiral field $w$ a very similar story holds. Its free field action is now given by,
\begin{eqnarray}
{\cal S}_{\mbox{\footnotesize twisted chiral}}=-\int d^2 \sigma d^4 \theta\, w \bar w\,,
\end{eqnarray}
and gauging it results in,
\begin{eqnarray}
{\cal S}_{\mbox{\footnotesize twisted chiral}}=-\int d^2 \sigma d^4 \theta\, w\, e^{\frac 1 2\, \bar\ID_-\big(F^=\ID_-}\bar w\,, 
\end{eqnarray}
which is invariant under the gauge transformations,
\begin{eqnarray}
&& w \rightarrow w'= e^{{\cal O}(\varepsilon^=)}w\,,\qquad
\bar w \rightarrow \bar w'= e^{{\cal O}(\bar\varepsilon^=)}\bar w\,, \nonumber\\
&&e^{i{\cal O}(\,F^=)} \rightarrow e^{i{\cal O}( F'^=)}= e^{{\cal O}(\bar\varepsilon^=)}e^{i{\cal O}(F^=)}e^{-{\cal O}(\varepsilon ^=)}\,.
\end{eqnarray}
with $\bar \varepsilon^= =\varepsilon^{=\dagger}$ and $\bar\ID_+ \varepsilon ^==\ID_+ \bar\varepsilon^==0$. In the same way as what was done for a chiral superfield one shows that this  describes a chiral boson satisfying $\bar \ID_\pm w=\ID_-w=0$ and $\ID_\pm\bar w=\bar\ID_-\bar w=0$ implying $\partial _=w=\partial _=\bar w=0$.

\subsection{Semi-chiral superfields}

The free field action for a semi-chiral multiplet is given by\footnote{The free action for the semi-chiral is by no means unique, any Legendre transform of the potential will do as well. See {\em e.g.} \cite{Sevrin:2011mc}.},
\begin{eqnarray}
&&{\cal S}_{\mbox{\footnotesize semi-chiral}}=\int d^2 \sigma d^4 \theta\, \big(\sqrt{2}\,(l\bar r+\bar l r)-
l \bar{l} - r \bar{r}
\big)\,,
\end{eqnarray}
and it is invariant under the superconformal transformations eq.~(\ref{suco222}) (where one uses the constraints eq.~(\ref{csch})).
The constraints eq.~(\ref{csch}) imply ,
\begin{eqnarray}
 \hat{D}_+ l =  i D_+ l  \ , \quad  \hat{D}_+ \bar l =  - i D_+ \bar l   \ , \quad  \hat{D}_- r   =  - i D_- r   \ , \quad  \hat{D}_- \bar r  =  - i D_- \bar r   \ ,  
\end{eqnarray}  
with which one passes to $N=(1,1)$ superspace,
\begin{eqnarray}
{\cal S}_{\mbox{\footnotesize semi-chiral}}&=&4\int d^2 \sigma d^2 \theta\, \Big(
D_+lD_-\bar l+D_+rD_-\bar r - \frac{1}{\sqrt 2}\,D_+lD_-\bar r- \frac{1}{\sqrt 2}\,D_+rD_-\bar l \nonumber\\
&&+ \frac{1}{2\sqrt 2}\big(\hat D_+r-i\sqrt 2 D_+l+iD_+r\big)
\big(\hat D_-\bar l+i\sqrt 2	D_-\bar r -iD_-\bar l\big)\nonumber\\
&&+ \frac{1}{2\sqrt 2}\big(\hat D_+\bar r+i\sqrt 2 D_+\bar l-iD_+\bar r\big)
\big(\hat D_- l-i\sqrt 2	D_-r+iD_- l\big)\,.\label{asc11}
\end{eqnarray}  
One sees that $\hat D_+r$, $\hat D_+\bar r$, $\hat D_-l$ and $\hat D_-\bar l$ are auxiliary fields 
and the last two lines in eq.~(\ref{asc11}) vanish when passing to a second order formalism. Making a
 coordinate transformation $r \rightarrow  r'=\sqrt 2\,\bar r-\bar l$ and $l \rightarrow  l'=l$ or equivalently 
$l \rightarrow l'=\sqrt 2\, \bar l-\bar r$ and $r \rightarrow r'=r$ brings the (second order) action to the customary diagonal 
form\footnote{In a second order formalism the constraints on the superfields assume the form 
$\hat D_\pm X=J_\pm D_\pm X$ where $J_+$ and $J_-$ are complex structures. For a model described 
in terms of semi-chiral super fields $\ker [J_+,J_-]=0$ holds and neither $J_+$ nor $J_-$ are diagonal. In \cite{Sevrin:1996jr} it was shown that given a generalized K\"ahler potential $V$, a coordinate transformation $r \rightarrow r'= \partial _l V$ and $\bar r \rightarrow \bar r'= \partial _{\bar l} V$ ($l \rightarrow l'= \partial _r V$ and $\bar l \rightarrow \bar l'= \partial _{\bar r} V$) while keeping the other coordinates fixed,
diagonalizes $J_+$ ($J_-$).},
\begin{eqnarray}
{\cal S}_{\mbox{\footnotesize semi-chiral}}&=&2\int d^2 \sigma d^2 \theta\, \Big(
D_+l'D_-\bar l'+D_+r'D_-\bar r'\big) \,.
\end{eqnarray}    

Using the results obtained in subsection 3.3 one can gauge this symmetry arriving at the action,
\begin{eqnarray}
{\cal S}_{\mbox{\footnotesize semi-chiral}}=\int d^2 \sigma d^4 \theta\, \big(\sqrt{2}\,(l\bar r+\bar l \,e^{+i {\cal O}(F^=)}r)-
l\, e^{-i {\cal O}(F^=)} \bar{l} - r\bar{r}  \label{gaugedsemichiral}
\big)\,,
\end{eqnarray}
which is invariant under the gauge transformations,
\begin{eqnarray}
&& l \rightarrow l'= e^{{\cal O}(\varepsilon^=)}l\,,\qquad
\bar l \rightarrow \bar l'= e^{{\cal O}(\bar\varepsilon^=)}\bar l\,, \nonumber\\
&& r \rightarrow r'= e^{{\cal O}(\varepsilon^=)}r\,,\qquad
\bar r \rightarrow \bar r'= e^{{\cal O}(\varepsilon^=)}\bar r\,, \nonumber\\
&&e^{i{\cal O}(\,F^=)} \rightarrow e^{i{\cal O}( F'^=)}= e^{{\cal O}(\bar\varepsilon^=)}e^{i{\cal O}(F^=)}e^{-{\cal O}(\varepsilon ^=)}\,.
\end{eqnarray}
where $\bar \varepsilon^= =\varepsilon^{=\dagger}$ and $\bar\ID_+ \varepsilon ^==\ID_+ \bar\varepsilon^==0$.

The most expedient way to understand the physical content  is by making the gauge choice given in eq.~(\ref{Fgc11})  and passing to  $N=(1,1)$ superspace. In this gauge the action eq.~(\ref{gaugedsemichiral})  simplifies to,
\begin{eqnarray}
{\cal S}_{\mbox{\footnotesize semi-chiral}}&=&\int d^2 \sigma d^4 \theta\, \big(\sqrt{2}\,(l\bar r
+\bar l r)-l \bar{l} - r \bar{r} \nonumber\\
&&
\qquad+\hat \theta^+h_{\pp\,+}(\sqrt 2\,\ID_-r\,\bar \ID_-\bar l+\ID_-\bar l\,\bar\ID_-l-\ID_-l\,\bar\ID_-\bar l\,)
\big)\,.
\end{eqnarray}
Passing to $N=(1,1)$ superspace one finds that the action eq.~(\ref{asc11}) gets modified to,

\begin{eqnarray}
&&{\cal S}_{\mbox{\footnotesize semi-chiral}}=4\int d^2 \sigma d^2 \theta\, \Big(
\nabla_+lD_-\bar l+\nabla_+rD_-\bar r - \frac{1}{\sqrt 2}\,\nabla_+lD_-\bar r- \frac{1}{\sqrt 2}\,\nabla_+rD_-\bar l \nonumber\\
&&+ \frac{1}{2\sqrt 2}\big(\hat D_+r-i\sqrt 2 \nabla_+l+i\nabla_+r-D_-h_{\pp\,+}D_-r-ih_{\pp\,+}\partial _=r\big)
\big(\hat D_-\bar l+i\sqrt 2	D_-\bar r -iD_-\bar l\,\big)\nonumber\\
&&+ \frac{1}{2\sqrt 2}\big(\hat D_+\bar r+i\sqrt 2 \nabla_+\bar l-i\nabla_+\bar r
-D_-h_{\pp\,+}D_-\bar r-ih_{\pp\,+}\partial _=\bar r
\big)
\big(\hat D_- l-i\sqrt 2	D_-r+iD_- l\big)\,,\nonumber\\
&&\label{ascgf11}
\end{eqnarray}  
where the covariant derivative $\nabla_+$ was defined in eq.~(\ref{covdev11}). Passing to the second order formalism -- which amounts to dropping the last two terms in eq.~(\ref{ascgf11}) -- and comparing the
result to eq.~(\ref{n11gauged}), we find that we indeed describe two complex (or four real) chiral scalars.

\section{Discussion and conclusions}

We have seen how to obtain both $N=(1,1)$ and $N=(2,2)$ PST-like formalism for chiral bosons in two-dimensions.  This is achieved by gauging chiral super-conformal transformations to give rise to firstly a Siegel-like formalism.  Adopting a Beltrami parametrisation for the gauge field gives rise to the PST treatment.  Finally one may choose to gauge fix and this reduces the theories to be of Floreanini-Jackiw type.  

We believe these results may prove to have utility when used in conjunction with the doubled (T-duality symmetric)  formalism of string theory 
\cite{Tseytlin:1990nb,Tseytlin:1990va,Hull:2004in}.  These approaches, in which chiral bosons play a central r\^ole, give a concrete proposal for the treatment of non-geometric or T-fold backgrounds at the level of the world sheet.  Although the full geometry appropriate for describing these theories is not yet fully understood their have been some hints of a very rich structure that shares many similarities with Hitchin's generalised geometry \cite{Hitchin:2004ut}.  Since sigma models in $N=(2,2)$ superspace provide an explicit local realisation of  generalised K\"ahler geomtry one might hope that a doubled formalism in   $N=(2,2)$ superspace will shed light on these novel geometric structures.  To make contact with the Doubled Field Theory formulation of supergravity one might then consider the conditions of conformal invariance of such a sigma model along the lines developed in \cite{Berman:2007xn}.

There are some subtle questions about chiral bosons in two-dimensions (and not just these $N=(2,2)$ supersymmetric ones) that are worth mentioning and which represent  important topics for further study.  First is the question of how these approaches can be extended to arbitrary genus Riemann surfaces as would be necessary for perturbative string theory applications.  The second is   the subtle question  of quantisation about which the historical literature is rather ambiguous and often conflicting.  Although we do not intend to present a resolution here we give a quick self contained review in the hope that it inspires further study of these issues. 

The most clear cut case is the Floreanini-Jackiw formalism \cite{Floreanini:1987as}.  Here quantisation can be readily performed, there are no gauge symmetries so nothing needs to be fixed.  This is a first class system with a second class constraint so the passage to the quantum theory can be achieved by means of Dirac brackets. As detailed in \cite{Sonnenschein:1988ug} one finds that the single particle Hilbert space correctly implements the chirality constraint and consists of a continuum of states with $E=k$ and with only positive (i.e. right moving)  $k\geq 0$ excitations.   It is in this formalism that some quantum questions of the doubled formalism have been addressed \cite{HackettJones:2006bp,Berman:2007xn}. 

The case of the Siegel formulation is rather more divisive.   One observation is that either by using Hamiltonian reduction as described in \cite{Faddeev:1988qp} or Dirac procedure as in \cite{Bernstein:1988zd} the Siegel formalism reduces to the Floreanini-Jackiw formalism and quantisation can proceed accordingly.  An alternative approach would be to treat the Siegel gauge symmetry directly using Faddeev-Popov \cite{Imbimbo:1987yt} and BRST methods \cite{Labastida:1987zy}. Doing so one encounters some puzzling results.  An immediate problem is that Siegel symmetry is anomalous. To see this  it very apposite to consider the
(non supersymmetric) Siegel action, 
\begin{eqnarray}
S= \int d^2\sigma \,\big(\partial_\pp \phi \, \partial_= \phi - h_{\pp\,\pp}\,\partial_= \phi\, \partial_=\phi \big) \,, 
\end{eqnarray}
 as a scalar field coupled to gravity in the light cone gauge;  the Lagrange multiplier, $h$, simply plays the r\^ole of the remaining component of the graviton.   Classically, this action is invariant under the gauge symmetries, 
 \begin{eqnarray}
\label{htrans}
&&\delta \phi = \epsilon^= \partial_= \phi \ , \quad \delta h_{\pp\,\pp}\, = \partial_\pp \epsilon^=   + \epsilon^= \partial_= h_{\pp\,\pp}\,  - (\partial_= \epsilon^=) h_{\pp\,\pp} \,.
\end{eqnarray} 
  
The anomalous Ward identity\footnote{In this particular case the central charge is $c=1$.}, 
\begin{eqnarray}
(\partial_\pp - 2 \partial_= h_{\pp\,\pp}\, - h_{\pp\,\pp}\, \partial_=) \frac{\delta \Gamma[h_{\pp\,\pp}\,] }{\delta h_{\pp\,\pp}\,} = \frac{c}{12 \pi} \partial_=^3 h_{\pp\,\pp}\,\,, 
\end{eqnarray} 
 provides a functional differential equation  which  can be exactly solved \cite{Schoutens:1991sp} to recover the {\emph{ non-local}} effective action \cite{Polyakov:1987zb}
\begin{eqnarray}
\label{Seff}
\Gamma^{\tiny{1-loop}} [h_{\pp\,\pp}\,] = \frac{c}{24 \pi} \int \partial_=^2 h_{\pp\,\pp}\, \frac{1}{\partial_\pp} \frac{1}{1 - h_{\pp\,\pp}\, \frac{\partial_=}{\partial_\pp} } \frac{1}{\partial_=}  \partial_=^2 h_{\pp\,\pp} \, ,
 \end{eqnarray} 
 which has an anomaly 
 \begin{eqnarray}
 \label{anom}
 \delta \Gamma[h_{\pp\,\pp}\,] =
 - \frac{c}{12 \pi}  \int d^2 \sigma  \epsilon^{=} \partial_=^3 h_{\pp\,\pp}\,   \, .
 \end{eqnarray}
 To counter  this anomaly and to restore the  symmetry at a quantum level, the authors of  \cite{Imbimbo:1987yt}  propose adding a Wess-Zumino term to the action
\begin{eqnarray}
S_{WZ}= \kappa \int d^2\sigma   h_{\pp\,\pp}\, \partial^2_= \phi \, . 
\end{eqnarray}  
This extra term explicitly violates the classical gauge symmetry however,   $\kappa$ may appropriately tuned (and including ghost loops where required),  such that this contribution cancels with that of  eq.~(\ref{anom}) leaving a quantum mechanically gauge invariant system.   Unfortunately this approach has difficulties when then coupled to world sheet gravity.  It was further argued   \cite{Henneaux:1987hz} that such an approach is fundamentally inapplicable because the first class constraint associated to the gauge symmetry of the Siegel action  is the square of a second class constraint.  
 
The quantum aspects of the PST formulation are less discussed, though given the wide use of the formalism rather important to understand.  In \cite{Pasti:1996vs} some classical arguments are presented as to why  the system may be well behaved quantum mechanically.  Firstly  it is observed that  first class constraint for the PST symmetry is {\em not} the square of a second class one.  Since the algebra of the constraints behaves like a Virasoro algebra one can guess that its quantum commutator may acquire a central charge which may be interpreted as the presence of an anomaly in the gauge symmetry.  In  \cite{Pasti:1996vs} it is argued that by appropriately modifying the definition of the constraint, without spoiling the property that it is classically first class, one can engineer and cancellation of the central charge in the quantum constraint algebra.   This argument was made at a rather heuristic level in  \cite{Pasti:1996vs}   and a step to making it more precise was made by Lechner in \cite{Lechner:1998ga} where it was demonstrated explicitly that upon coupling to gravity the only extra contributions to the expected  gravitational anomaly of a chiral boson  \cite{AlvarezGaume:1983ig} were trivial (i.e. removable by local finite counter terms).   One might ask whether this can be made precise without the extra complication of coupling to gravity. 
 
Using the above results we can directly write down the effective action for the PST formalism.   The PST action in this case is given by 
\begin{eqnarray}
S= \int d^2\sigma \,\big(\partial_\pp \phi\,  \partial_= \phi -  \frac{ \partial_\pp f }{\partial_= f }   \partial_= \phi \,\partial_=\phi\big)  \, , 
\end{eqnarray}
To pass from the Siegel formulation to the PST one invokes a Beltrami parametrisation for the metric in light cone gauge namely 
\begin{eqnarray}
\partial_\pp  f = h_{\pp\,\pp}\, \partial_= f \, . 
\end{eqnarray}
 Polyakov \cite{Polyakov:1987zb}   showed that under this transformation the non-local effective action eq.~(\ref{Seff}) becomes  local (albeit non-polynomial) and is given by\footnote{This equation corrects a typo in this equation (5) of \cite{Polyakov:1987zb}.}
\begin{eqnarray}
\Gamma^{\tiny{1-loop}} [f] = \frac{c}{24\pi}  \int d^2 \sigma\, \left[ \frac{\partial_=^2 f \partial_{\pp  }\partial_= f }{ (\partial_= f)^2 } - \frac{(\partial^2_= f)^2 \partial_\pp f}{ (\partial_= f)^3}  \right] \, . 
\end{eqnarray}
This result gives the effective action for the PST formulation.  It has the same non-vanishing variation under the gauge symmetries as  eq.~(\ref{Seff}) so it is not immediately clear why this should not be considered an anomaly.   One possible resolution is that the effective action in this case is local (it has no $\Box^{-1}$ term).  Then one might be tempted to use the freedom to subtract local counter terms to remove it and declare the anomaly to be trivial.  But this last step needs considerable refinement and care. Indeed, from the outset the sort of action used in the PST formalism is non-standard; it is non-polynomial with derivatives entering in both numerator and denominator. Thus the formal treatment of renormalisation needs to be carefully considered.  Such a treatment is beyond the scope of this note but represents an interesting area for study in its own right.

\acknowledgments

It is a pleasure to thank Frank Ferrari, Dmitri Sorokin, Arkady Tseytlin and especially Marc Henneaux for enlightening 
discussions and correspondence. The authors are supported in part by the Belgian Federal Science Policy Office
through the Interuniversity Attraction Pole P7/37, and in part by the
``FWO-Vlaanderen'' through the project G.0114.10N and by the Vrije Universiteit Brussel through the Strategic Research Program ``High-Energy Physics''. DT is an FWO postdoc. 

\appendix

\section{Conventions, notations and identities}\label{app conv}
We denote the worldsheet coordinates by $ \tau,\sigma \in\IR$, and the worldsheet light-cone coordinates are
defined by,
\begin{eqnarray}
\sigma ^\pp= \tau + \sigma ,\qquad \sigma ^== \tau - \sigma\, .\label{App1}
\end{eqnarray}
The $N=(1,1)$ (real) fermionic coordinates are denoted by $ \theta ^+$ and $ \theta ^-$ and the
corresponding derivatives satisfy,
\begin{eqnarray}
D_+^2= - \frac{i}{2}\, \partial _\pp \,,\qquad D_-^2=- \frac{i}{2}\, \partial _= \,,
\qquad \{D_+,D_-\}=0\,.\label{App2}
\end{eqnarray}
The $N=(1,1)$ integration measure is explicitely given by,
\begin{eqnarray}
\int d^ 2 \sigma \,d^2 \theta=\int d^2 \sigma\,D_+D_- =2\int d\tau \,d \sigma \,D_+D_-\,.
\end{eqnarray}
Passing from $N=(1,1)$ to $ N=(2,2)$ superspace requires
the introduction of two more real fermionic coordinates $ \hat \theta ^+$ and $ \hat \theta ^-$
where the corresponding fermionic derivatives satisfy,
\begin{eqnarray}
\hat D_+^2= - \frac{i}{2} \,\partial _\pp \,,\qquad \hat D_-^2=- \frac{i}{2} \,\partial _= \,,
\end{eqnarray}
and again all other -- except for (\ref{App2}) -- (anti-)commutators   vanish.
The $N=(2,2)$ integration measure is,
\begin{eqnarray}
\int d^2 \sigma \,d^2 \theta \, d^2 \hat \theta =2
\int d \tau\, d \sigma \,D_+D_-\, \hat D_+ \hat D_-\,.
\end{eqnarray}
Regularly a complex basis is used,
\begin{eqnarray}
\ID_\pm\equiv \hat D_\pm+i\, D_\pm,\qquad
\bar \ID_\pm\equiv\hat D_\pm-i\,D_\pm\,,
\end{eqnarray}
which satisfy,
\begin{eqnarray}
\{\ID_+,\bar \ID_+\}= -2i\, \partial _\pp\,,\qquad
\{\ID_-,\bar \ID_-\}= -2i\, \partial _=\,,
\end{eqnarray}
and all other anti-commutators   vanish.


\section{$N=1$ superconformal transformations}
An $N=(1,1)$ chiral coordinate transformation $\sigma^= \rightarrow  \sigma'{}^=( \sigma ^=, \theta^-)$,  $\theta^- \rightarrow  \theta '{}^-( \sigma ^=, \theta^-)$ is said to be superconformal if the fermionic derivative transforms covariantly, $D_-=D_- \theta' {}^-D'_-$, which holds if \cite{Schoutens:1988ig},
\begin{eqnarray}
D_- \sigma'^=+\frac i 2 \, \theta'^- D_- \theta '^-=0\,.\label{succond1}
\end{eqnarray}
This in turn implies,
\begin{eqnarray}
\partial _= \sigma'^=-\frac i 2 \, \theta'^- \partial _= \theta'^-= \big(D_- \theta '^-\big)^2\,.
\end{eqnarray}
For infinitesimal transformations --- $ \sigma'^== \sigma ^=+ \delta \sigma^=$ and 
$ \theta'^-= \theta ^-+ \delta \theta^-$ ---
this condition can be solved in terms of a single superfield $\varepsilon^=$,
\begin{eqnarray}
\delta \sigma^=&=& \varepsilon^=-\frac 1 2 \,\theta^-D_- \varepsilon ^= \,, \nonumber\\
\delta \theta^-&=&i\, D_- \varepsilon ^=\,.
\end{eqnarray}
Before considering finite transformations we introduce the operator\footnote{Note that for an anti-commuting field $\psi$ we get 
${\cal O}(\psi)=\psi\,\partial _=-i \, D_-\psi  D_-\,$.} ${\cal O}(A^=)$,
\begin{eqnarray}
{\cal O}(A^=)\equiv A^= \partial _=+ i  \, D_-A^=D_-\,,
\end{eqnarray}
which closes under commutation,
\begin{eqnarray}
[ {\cal O}(A^=),{\cal O}(B^=)]= {\cal O}\big(
A^= \partial _=B^=- \partial _=A^= B^=+iD_-A^=D_-B^=
\big)\,.
\end{eqnarray}
With this we can construct the finite transformations,
\begin{eqnarray}
\theta'^-=e^{ {\cal O}( \varepsilon^=)}\, \theta^-\,e^{ -{\cal O}( \varepsilon^=)}\,,\qquad
\sigma'^==e^{ {\cal O}( \varepsilon^=)}\, \sigma^=\,e^{ -{\cal O}( \varepsilon^=)}\,.\label{bb6}
\end{eqnarray} 
We now show that the transformations eq.~(\ref{bb6})  satisfy eq.~(\ref{succond1}). Introduce $D'_-\,$,
\begin{eqnarray}
D'_-=e^{{\cal O}( \varepsilon^=)}D_-e^{-{\cal O}( \varepsilon^=)}\,.\label{bb7}
\end{eqnarray}
Using eq.~(\ref{bb6}) one shows that this indeed satisfies $D'_- \theta'^-=1- \theta '^-D'_-\,$. Using the fact that $[ {\cal O}(\varepsilon^=), f\,D_-]\propto D_-$ where $f$ is an arbitrary function, one shows with the defining equation eq.~(\ref{bb7}) that $D'_-=K\,D_-$ where $K$ is some function of $\varepsilon^=$ and its derivatives. From $1=(D'_-\theta'^-)=K(D_- \theta '^-)$ one determines $K$,
\begin{eqnarray}
D'_-= \frac{1}{D_- \theta'^-}\, D_-\,,\label{bb8}
\end{eqnarray}
showing that the fermionic derivative indeed transforms covariantly. The rest of the proof is then straightforward. Using eq.~(\ref{bb8}) we find,
\begin{eqnarray}
D'_- \sigma'^== \frac{1}{D_- \theta'^-}\, D_- \sigma '^=\,.\label{bb9}
\end{eqnarray} 
On the other hand we can calculate $D'_- \sigma'^=$ usings eqs.~(\ref{bb7}) and (\ref{bb6}),
\begin{eqnarray}
D'_- \sigma'^== -\frac i 2 \, \theta'^-\,.\label{bb10}
\end{eqnarray}
The combination of eq.~(\ref{bb9}) with eq.~(\ref{bb10}) gives eq.~(\ref{succond1}).   

In the whole discussion we tacitly assumed that $D_+ \varepsilon^==0$. However, even when this is not true, the transformations in eq.~(\ref{bb6})
still satisfy the condition eq.~(\ref{succond1})!  
 
\section{$N=2$ superconformal transformations}
The extension to the $N=(2,2)$ superconformal case is rather straightforward. A chiral $N=(2,2)$ coordinate transformation,
\begin{eqnarray}
\sigma^= \rightarrow \sigma'^=( \sigma ^=, \theta^-, \bar \theta ^-)\,,\qquad
\theta^- \rightarrow \theta'^-( \sigma ^=, \theta^-, \bar \theta ^-)\,,\qquad
\bar\theta^- \rightarrow \bar\theta'^-( \sigma ^=, \theta^-, \bar \theta ^-)\,,\label{B9}
\end{eqnarray}
is superconformal if the fermionic derivatives transform covariantly.
\begin{eqnarray}
\ID_-=\ID_- \theta'^-\,\ID'_- +\ID_- \bar\theta'^-\,\bar\ID'_-\,,\qquad
\bar\ID_-=\bar\ID_- \theta'^-\,\ID'_- +\bar\ID_- \bar\theta'^-\,\bar\ID'_-\,.
\label{C2}
\end{eqnarray}
This is true provided,
\begin{eqnarray}
\ID_- \sigma'^=+i\, \theta'^-\ID_-\bar \theta '^-
+i\, \bar\theta'^-\ID_-\theta '^-=
\bar\ID_- \sigma'^=+i\, \theta'^-\bar\ID_-\bar \theta '^-
+i\, \bar\theta'^-\bar\ID_-\theta '^-=0\,,
\end{eqnarray}
holds. The condition above implies several integrabiltity conditions,
\begin{eqnarray}
&&\ID_- \theta'^-\ID_-\bar \theta '^-=\bar\ID_- \theta'^-\bar\ID_-\bar \theta '^-=0\,, \nonumber\\
&& \partial _= \sigma'^=-i \,\theta'^- \partial _= \bar \theta'{}^--i \,\bar\theta'{}^- \partial _=  \theta'{}^-=
\ID_- \theta'^-\bar\ID_-\bar \theta '^-+\bar\ID_- \theta'^-\ID_-\bar \theta '^-
\,.
\end{eqnarray}
The first set of conditions are solved by imposing,
\begin{eqnarray}
\bar\ID_- \theta'^-=\ID_-\bar \theta '{}^-=0\,,
\end{eqnarray}
which results in invertible coordinate transformations containing the identity transformation. Summarizing, an $N=(2,2)$ superconformal transformation
eq.~(\ref{B9})  satisfies,
\begin{eqnarray}
&&\ID_- \sigma'^=
+i\, \bar\theta'^-\ID_-\theta '^-=
\bar\ID_- \sigma'^=+i\, \theta'^-\bar\ID_-\bar \theta '^-=0\,, \nonumber\\
&&\bar\ID_- \theta'^-=\ID_-\bar \theta '{}^-=0\,.\label{B13} \label{C6}
\end{eqnarray}

For an infinitesimal superconformal transformation, $\sigma^= \rightarrow  \sigma'^== \sigma ^=+ \delta \sigma^=$,
$\theta^- \rightarrow \theta'^-= \theta ^-+ \delta \theta^-$, $\bar\theta^- \rightarrow \bar\theta'^-= \bar\theta ^-+ \delta \bar\theta^-$,
one solves eq.~(\ref{B13}) in terms of a single superfield $\varepsilon^=$,
\begin{eqnarray}
\delta \sigma^=&=& \varepsilon^=-\frac 1 2 \, \theta^-\,\ID_- \varepsilon^=-\frac 1 2 \, \bar\theta^-\,\bar\ID_- \varepsilon^=\,,\nonumber\\
\delta \theta^-&=&\frac i 2 \bar\ID_- \varepsilon^=\,,\nonumber\\
\delta \bar\theta^-&=&\frac i 2 \ID_- \varepsilon^=\,.
\end{eqnarray} 

Before proceeding we introduce the notation,
\begin{eqnarray}
{\cal O}\big(A\big)&=&\frac i 2\, \ID_-\big(A\,\bar\ID_-\,+\frac i 2\, \bar\ID_-\big(A\,\ID_-\nonumber\\
&=& A\,\partial _=+\frac i 2\,\big( \ID_-A\big)\,\bar\ID_-+\frac i 2 \, \big(\bar \ID_-A\big)\,\ID_-\, .
\end{eqnarray}
A useful property in what follows is the closure of this operator given by,
\begin{eqnarray}
\big[
{\cal O}(A), {\cal O}(B)\big]= {\cal O}\big(
A \,\partial _= B +\frac i 2 \, \bar\ID_-A\,\ID_-B+\frac i 2 \, \ID_-A\,\bar\ID_-B- \partial _= A \,B\,  
\big) \, .
\end{eqnarray}
With this we can write the finite the finite transformations as,
\begin{eqnarray}
\sigma'^=&=&e^{ {\cal O}( \varepsilon^=)}\, \sigma^=\,e^{ -{\cal O}( \varepsilon^=)}\,,\nonumber\\
\theta'^-&=&e^{ {\cal O}( \varepsilon^=)}\, \theta^-\,e^{ -{\cal O}( \varepsilon^=)}\,, \nonumber\\
\bar\theta'^-&=&e^{ {\cal O}( \varepsilon^=)}\,\bar \theta^-\,e^{ -{\cal O}( \varepsilon^=)}\,.
\label{sprime}
\end{eqnarray}
We may repeat similar steps as in the $N=(1,1)$ case to show that these finite transformations do indeed obey eq.~(\ref{C6}).  Since $[ {\cal O}, f \ID_-] \propto \ID_- $ for an arbitrary function $f$,  one can conclude (for instance using a Baker-Campbell-Hausdorff expansion) that
\begin{eqnarray}
\ID_-^\prime =  e^{ {\cal O}( \varepsilon^=) }\ID_-   e^{- {\cal O}( \varepsilon^=)}= K \ID_-  \ . 
\end{eqnarray}
By observing that  $\ID_-^\prime \theta'  = 1$ one determines $K = \frac{1}{ \ID_- \theta'^-}$.  Then one can see at once that
\begin{eqnarray}
\frac{1}{ \ID_- \theta'^-} \ID_- \bar\theta'^-   =  \ID_-^\prime  \bar\theta'^-  = 0 \ .
\end{eqnarray}
A similar result holds for the barred derivative and so the derivatives transform covariantly in accordance with eq.~(\ref{C2}) and the second line of eq.~(\ref{C6}).  To obtain the first line of eq.~(\ref{C6}) one notes that $\ID_- \sigma^= = -i \bar\theta^-$ so that 
\begin{eqnarray}
\frac{1}{ \ID_- \theta'^-} \ID_- \sigma'^=   =  \ID_-^\prime  \sigma'^= =  - i \bar\theta'^- \ ,
\end{eqnarray}
and a similar relation for barred derivatives. 

Now we wish to show that the gauge covariant derivative
\begin{eqnarray}
\nabla_+  = e^{{\cal O} (V^=)}     \ID_+ e^{-{ \cal O}(V^=)}  = \ID_+ - h_{\pp\, +}   \partial_= + \frac{i}{2} \bar\ID_-   h_{\pp\, +}   \ID_- + \frac{i}{2}  \ID_-   h_{\pp\, +}  \bar \ID_-  \, , 
\end{eqnarray}
may be expressed in a Beltrami parametrisation by 
\begin{eqnarray}
h_{\pp\, +}  = \frac{\ID_+ g^= + i \psi^- \ID_+ \bar\psi^- + i \bar\psi^- \ID_+ \psi^- }{\ID_-\psi^- \bar\ID_-\bar{\psi}^-}
\label{belt2}
\end{eqnarray}
with
\begin{eqnarray}
g^=&=&e^{ {\cal O}(V^=)}\, \sigma^=\,e^{ -{\cal O}( V^=)}\,,\nonumber\\
\psi^-&=&e^{ {\cal O}( V^=)}\, \theta^-\,e^{ -{\cal O}( V^=)}\,, \nonumber\\
\bar\psi^-&=&e^{ {\cal O}(V^=)}\,\bar \theta^-\,e^{ -{\cal O}(V^=)}\,. 
\label{belt1}
\end{eqnarray}
To see this, note that
\begin{eqnarray}
\ID_+ g^= &=& ( \ID_+ e^{ {\cal O}(V^=)})   e^{-  {\cal O}(V^=)} g^=   +  g^=  e^{ {\cal O}(V^=)} (  \ID_+  e^{-  {\cal O}(V^=)}  ) \nonumber \\
&=&    h_{\pp\, +}   \partial_=  g^= -  \frac{i}{2} \bar\ID_-   h_{\pp\, +}   \ID_- g^=  + \frac{i}{2}  \ID_-   h_{\pp\, +}  \bar \ID_-  g^= \, . 
\end{eqnarray}
It then follows that
\begin{eqnarray}
&& \ID_+ g^= + i \psi^- \ID_+ \bar\psi^- + i \bar\psi^- \ID_+ \psi^- =  h_{\pp\, +} ( \partial_= g^= - i \bar \psi^- \partial_= \psi^- -  i \psi^- \partial_= \bar\psi^- ) \nonumber \\
 && \qquad\qquad   - \frac{i}{2} \bar \ID_- h_{\pp\,+} ( \ID_-g^= + i \bar{\psi}^- \ID_- \psi^-)   - \frac{i}{2} \ID_- h_{\pp\,+} ( \bar  \ID_-g^= + i  \psi ^- \ID_- \bar \psi^-)  \, . 
\end{eqnarray}
The fields defined in eq.~(\ref{belt1}) can be thought of as being obtained from an action of a superconformal transformation with parameter $V^=$ on the coordinates of superspace (c.f. eq.~(\ref{sprime})).  Hence they obey constraints of the form  eq.~(\ref{C6})  and as a consequence eq.~(\ref{belt2}) follows directly.


\end{document}